\documentclass{ws-ijmpb}

\newcommand{\insertwidegraphics}[2]{
	\begin{figure}
	\begin{center}
	\resizebox{\columnwidth}{!}{\includegraphics{#1}}
	\caption{#2}
	\end{center}
	\end{figure}
}

\begin{document}

\markboth{Akakii Melikidze}
{Spectral Function}

\catchline{}{}{}{}{}

\title{Effects of Quantum Hall Edge Reconstruction on Momenum-Resolved
	Tunneling}

\author{Akakii Melikidze and \underline{Kun Yang}}
\address{National High Magnetic Field Laboratory,\\
		1800 E. Paul Dirac Dr.,\\
		Tallahassee, FL 32310
	\footnote{Email correspondence should be sent to:
	kunyang@magnet.fsu.edu}}

\maketitle

\begin{history}
\received{DAY MONTH YEAR}
\revised{DAY MONTH YEAR}
\end{history}

\begin{abstract}

During the reconstruction of the edge of a quantum Hall liquid,
Coulomb interaction energy is lowered through the change in
the structure of the edge. We use theory developed earlier by one of
the authors [K. Yang, Phys. Rev. Lett. {\bf 91}, 036802 (2003)]
to calculate the electron spectral functions of a reconstructed edge,
and study the consequences of the edge
reconstruction for the momentum-resolved tunneling into the edge. It
is found that additional excitation modes that appear after the
reconstruction produce distinct features in the energy and momentum
dependence of the spectral function, which can be used to detect the
presence of edge reconstruction.

\end{abstract}



The paradigm of the Quantum Hall effect (QHE) edge physics is based on an
argument due to Wen,~\cite{Wen} according to which the low-energy edge
excitations are described by a chiral Luttinger liquid (CLL) theory.
One attractive feature of this theory is that due to the chirality,
the interaction parameter of the CLL is often tied to the robust topological
properties
of the bulk and is independent of the details of electron interaction and
edge confining potential; studying the physics at the edge thus offers
an important probe of the bulk physics. It turns out, however, that
the CLL ground state may not always be
stable.~\cite{MacDonald-Yang-Johnson,Chamon-Wen} On the microscopic level,
the instability is driven by Coulomb interactions and leads to
the change of the structure of the edge. This effect has been termed
``edge reconstruction''. One of its manifestations is the appearance of
new low-energy excitations of the edge not present in the original CLL
theory.

Previous numerical studies~\cite{Wan-Yang-Rezayi,Wan-Rezayi-Yang}
have suggested that the phenomenon of edge
reconstruction can be understood as an instability of the original
edge mode described by the CLL theory.
This instability occurs as a result of increasing curvature of the  
edge spectrum as the edge confining potential softens. The spectrum curves
down at high values of momenta until it touches zero at the transition point.
This signals an instability of the ground state as
the edge excitations begin to condense at finite momentum $k_0$.
Such condensation implies the appearance
of a bump in the electron density localized in real space at
$y_0 = l^2 k_0$, where $y_0$ is the coordinate normal to the edge and
$l$ is the magnetic length. These condensed excitations form
a superfluid (with power-law correlation) which possesses a
neutral ``sound'' mode that can propagate in both directions.


Following Ref.~6, we introduce two slowly varying fields
$\phi_1$ and $\varphi$ which describe the original charged edge mode
and the pair of the two new neutral modes, respectively. The total
action of the reconstructed edge of the FQHE at filling fraction
$\nu=1/m$ is: $S=S_1+S_\varphi+S_{\rm int}$, where
\begin{eqnarray}
\label{Action1}
S_1 &=& \frac{m}{4\pi}\int dt\,dx\,\left[\partial_t\phi_1\partial_x\phi_1
	-v(\partial_x\phi_1)^2\right],\\
\label{Action2}
S_\varphi &=& \frac{K_1}{2}\int dt\,dx\,
	\left[\frac{1}{v_\varphi}(\partial_t\varphi)^2
		-v_\varphi(\partial_x\varphi)^2\right],\\
\label{Action3}
S_{\rm int} &=& -K_2\int dt\,dx\, (\partial_t\varphi)(\partial_x\phi_1).
\end{eqnarray}
Here,
$K_1 \sim K_2\sim 1$, $v\gg v_\varphi$.
The latter inequality comes from the fact that the Coulomb interaction
boosts the velocity of the charged mode, but not those of neutral ones.

The action in Eqs.~(\ref{Action1},\ref{Action2},\ref{Action3})
describes the simplest situation
where edge excitation condensed in the vicinity of a single point $k_0$.
Such condensation may also occur at multiple points,
producing a set of pairs of additional neutral modes.
We shall comment on this below.

The theory~\cite{Yang} also
provides an explicit expression for the electron operator:
\begin{eqnarray}
\Psi = \exp\left\{im\phi_1 + \rho\cos(k_0x+\varphi)\right\}.
\end{eqnarray}
The constant $\rho$ is proportional to the density of
the new condensate and thus rises from zero at the reconstruction
transition.
We begin by evaluating the Green's function in real space and imaginary-time
representation: $g(x,\tau)=-i\langle \Psi(0,0)\Psi^\dag(x,\tau)\rangle$.
To this end, we write:
\begin{eqnarray}
\label{Expansion}
\Psi &=& e^{im\phi_1}\sum\limits_{n=-\infty}^{\infty}
	c_n e^{in(k_0x+\varphi)},
\end{eqnarray}
In the long wave length limit, the dominant contribution to $g(x,\tau)$
will come from the term with $n=0$ in the above series. Therefore,
we only retain this term in what follows; the effect of the omitted
terms will be commented upon in the discussion. Since the action of
system is Gaussian, the evaluation of the Green's function is now
straightforward.
One may exploit the limits $v_\varphi/v\ll 1$, $x,\tau\to\infty$ to
obtain:
\begin{eqnarray}
\label{Greens_function}
g(x,\tau) \propto \prod\limits_{i=1}^{3}\frac{1}{(x+iv_i\tau)^{\alpha_i}}.
\end{eqnarray}
Here $v_1$, $v_2$ and $v_3$ are the velocities of the three modes:
$v_1\approx v(1-2\beta)$, $v_{2,3}\approx \pm v_\varphi(1+\beta)$,
where $\beta = (\pi K_2^2/mK_1)(v_\varphi/v)$ is small.
To second order in $v_\varphi/v$, the exponents are:
$\alpha_1\approx m$, $\alpha_{2,3}\approx (4\pi K_2^2/K_1)(v_\varphi/v)^2$.
The sum of these three exponents describes local electron tunneling
into the reconstructed edge and has been obtained earlier.~\cite{Yang}

The spectral function is obtained from a Fourier transform of
Eq.~(\ref{Greens_function}).
A typical plot of the spectral function is shown in Fig.~1.a.
As a result of edge reconstruction, some
of the spectral weight is shifted away from the original $\delta$-function
singularity at $\omega=vq$ (its new position $v_1q$ is itself slightly
renormalized). An extra pair of singularities appear at
$\omega=v_{2,3} q$; these singularities correspond to the new
neutral modes. An important feature of the spectral function  is a
finite amount of
weight at $\omega<0$. This is possible only due to the fact that,
after the reconstruction, an excitation mode appears that propagates
in the direction opposite to the direction of the original edge
mode.


\insertwidegraphics{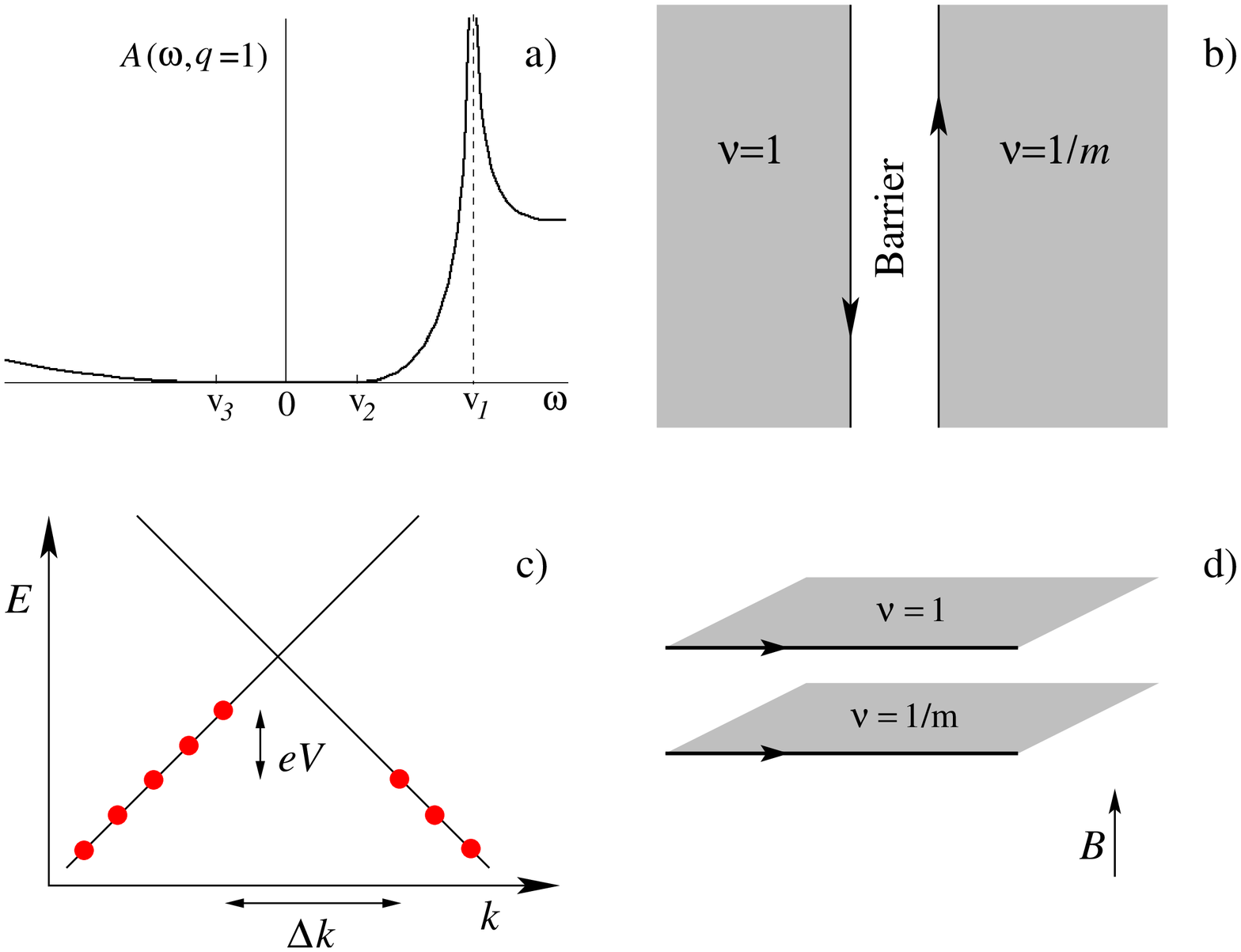}{a) The electron
spectral function in the reconstructed edge has singularities on
the lines $\omega=v_iq$ that correspond to the edge excitation modes.
The singularity at $\omega=v_1q$ is a divergence, and is the remnant of
the $\delta$-peak in the unreconstructed edge. The spectral function
vanishes for $v_3q<\omega<v_2q$ (e.g. for $q>0$) due to
kinematic constrains. Shown is the case of the filling fraction $\nu=1/3$.
b) The setup for the counter-propagating edge modes consists of $\nu=1$
and reconstructed $\nu=1/m$ states separated by a barrier which is uniform
along the edges. Magnetic field is pointing into the page.
c) The structure of the energy dispersion in the vicinity of
the barrier for the case of counter-propagating edge modes
(see text for details).
d) The double-layer structure with the co-propagating edge modes.}

The signatures of edge reconstruction in the spectral function are best
observed in the so-called momentum-resolved tunneling
experiments.~\cite{Kang,Grayson} In these experiments, the electron tunneling
into the edge occurs across
a barrier which is extended and homogeneous along the edge of the
FQHE system, and so the electron's momentum along the edge is conserved.
First, we consider the simplest possible case of tunneling
from the (un-reconstructed) edge of the QHE at the filling fraction
$\nu=1$.
The setup and the structure of energy levels under these conditions
are shown in Figs.~1.b,c.
The energy $E$ of the states is plotted as a function of
momentum $k$ along the two parallel edges.
The two straight lines in Fig.~1.c correspond to single electron 
dispersion on the two sides of the barrier.
We shall treat tunneling
using Fermi golden rule. Finally, we neglect the interaction between
the edge modes on the opposite sides of the barrier, while the intra-edge
interactions are taken into account by strong renormalization of the
(charged) mode velocities.


There are two parameters that one can control: bias voltage $V$ and 
magnetic field $B$;
$eV$ sets the difference between the
Fermi energies on the two sides of the barrier, while the difference
between the two Fermi momenta $\Delta k \propto B-B_0$, where $B_0$ is
the strength of the magnetic field at which, in the absence of bias, the
Fermi energy lies exactly at the dispersion lines crossing point.
Below, we shall rescale $B$ so that $\Delta k = B-B_0$.

Within our approximations, the tunneling current is:
\begin{eqnarray}
\label{Fermi}
I(V,B) &\propto& \int A_1(\omega_1,q_1)A_2(\omega_2,q_2)
	[f(\omega_1)-f(\omega_2)]\nonumber\\
	&&\times\delta(eV+\omega_1-\omega_2)\delta(B-B_0-q_1-q_2).
\end{eqnarray}
Here, $A_1=\delta(\omega-v_Fq)$ is the spectral function of the $\nu=1$ edge
with the Fermi velocity $v_F$, and $A_2$ is the spectral function of the
reconstruted edge.
$f(\omega)$ is the Fermi distribution
step function.

\insertwidegraphics{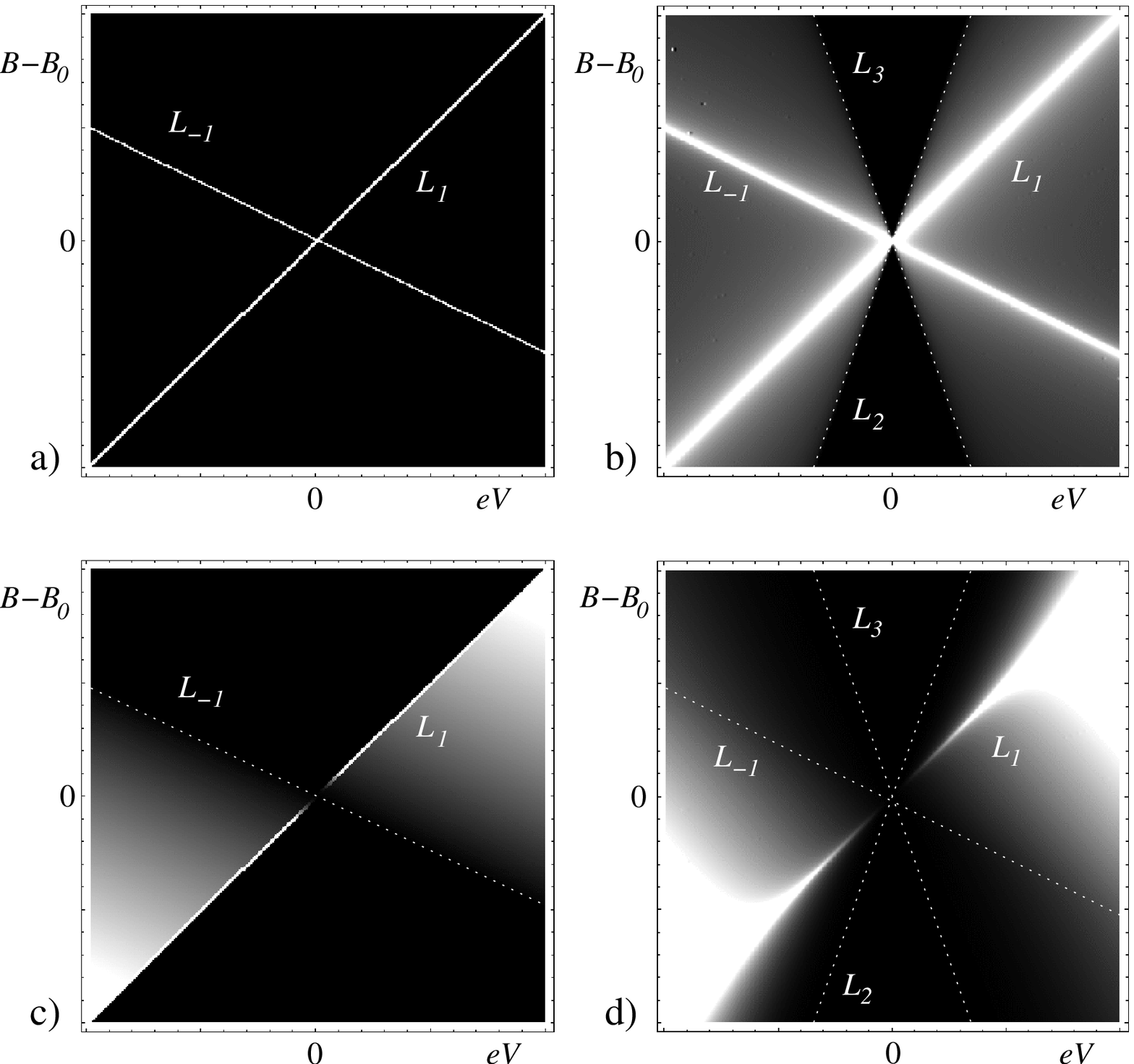}{Differential conductance $dI/dV$
as a function of bias voltage $V$ and magnetic field $B$
for momentum-resolved tunneling from a counter-propagating
unreconstructed $\nu=1$ edge to:
a) unreconstructed $\nu=1$ edge; b) reconstructed $\nu=1$ edge;
c) unreconstructed $\nu=1/3$ edge; d) reconstructed $\nu=1/3$ edge.
Plotted is the derivative of the result in Eq.~(\ref{Fermi}). Parameters
are: $v_1=1$, $v_{2,3}=\pm 1/3$, $v_F=2$, $\alpha_1=m=\nu^{-1}$,
$\alpha_{2,3}=0.2$. The dashed lines are guides to the eye and are defined
as $L_{1,2,3}=\{eV=v_{1,2,3}(B-B_0)\}$, $L_{-1}=\{eV=-v_F(B-B_0)\}$.
See text for details.}

The result of the numerical evaluation
of the differential conductance $dI/dV$ as a function of $V$ and $B-B_0$
is shown in Fig.~2. In general, the plot of
$dI(V,B)/dV$ is very similar, but not identical to $A(\omega,q)$.
In particular, both have 3 lines of singularities (marked by letters
$L_{1,2,3}$ on the figure), one of them being a divergence. These lines
correspond to the three edge excitation modes $\omega=v_iq$. Moreover,
in the bow-tie region between $L_2$ and $L_3$ the differential conductance
(as well as the current itself) is exactly zero. The reason for this is
purely kinematic: In the region below the dispersion line
of the slowest excitation in the system, one cannot satisfy the conservation
of energy and momentum in tunneling.

The general structure
of the spectral weight transfer provides a clear indication
of the edge reconstruction. In particular, in the bow-tie regions between
$L_1$ and $L_2$, and between $L_3$ and $L_{-1}$, $dI/dV$ is zero before
reconstruction and finite after. We note, that the rise of $dI/dV$
in the region between $L_3$ and $L_{-1}$ is due to the appearance of
a (neutral) edge mode that propagates in the direction opposite
to the direction of the original edge mode.

Most of the results discussed above are also valid for the double-layer
setup with co-propagating edge modes, shown in Fig.~1.d.
The differential conductance in this case is plotted in Fig.~3.
There is one crucial
difference, however: In the co-propagating case there is a region
(along $L_{-1}$) of negative differential conductance; reconstruction is
expected to progressively wash this region out. 

Although we have concentrated on the case of edge
reconstruction described by a single point $k_0$, the preceding discussion
remains generally valid for multiple edge reconstruction as well.
In that case, new lines of singularities appear, while $L_2$ and $L_3$
correspond to the two slowest excitation modes.

\insertwidegraphics{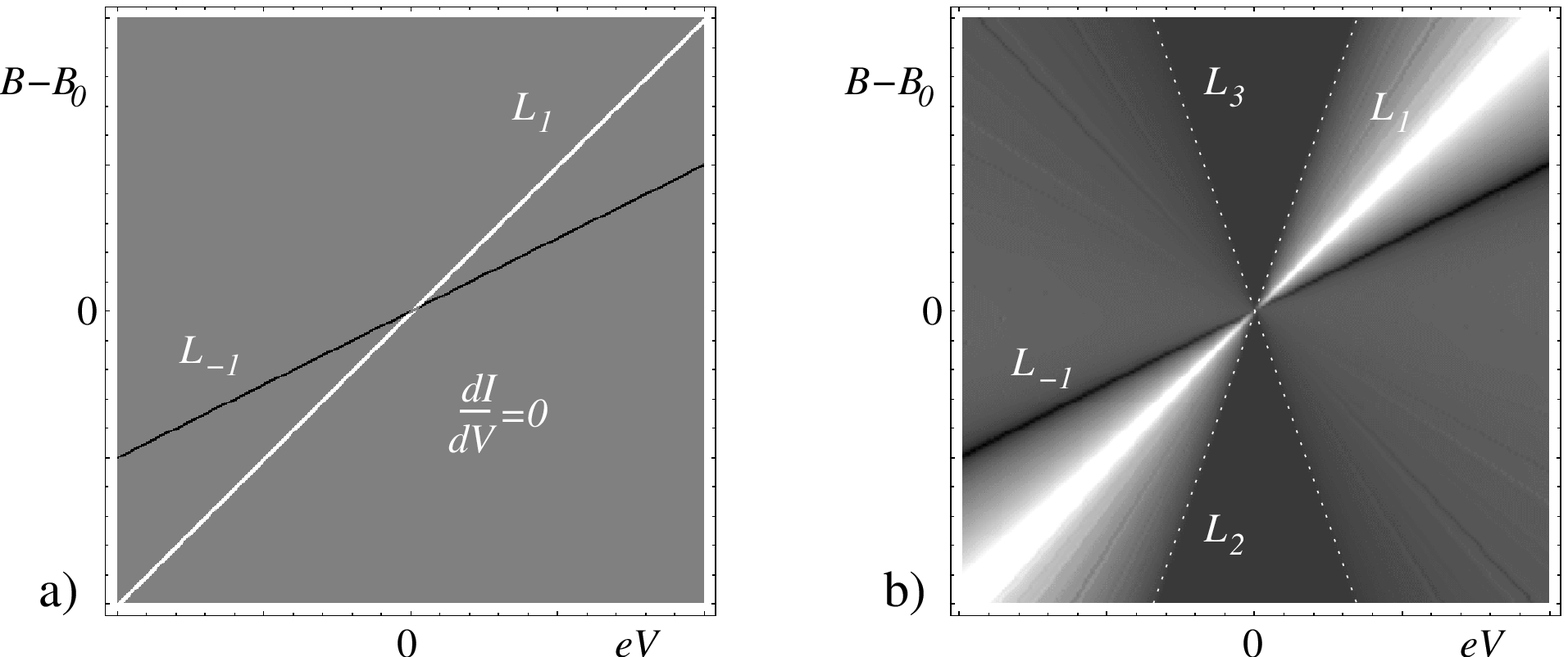}{a) The $dI/dV$ for tunneling from
unreconstructed $\nu=1$ to unreconstructed $\nu=1$ edge in the
co-propagating case. $dI/dV=0$ everywhere except: $dI/dV>0$ on $L_1$,
$dI/dV<0$ on $L_{-1}$.
b) The $dI/dV$ for tunneling from unreconstructed
$\nu=1$ to reconstructed $\nu=1$ edge in the co-propagating case.
$dI/dV=0$ in the region between $L_2$ and $L_3$.
Notation and parameters are the same as in Fig.~2,
except $v_F=-2$ is used for this plot.
Note that $dI/dV<0$ on $L_{-1}$.}

Finally, we would like to comment on the role of the omitted $n\neq 0$
terms in Eq.~(\ref{Expansion}). Apart from the factors $\exp ink_0x$, which
trivially shift the momentum argument of the electron spectral function
by $nk_0$, all terms with $n \neq 0$ are qualitatively similar to the
$n=0$ term. Each of them produces a contribution to the Green's function
which is of the from Eq.~(\ref{Greens_function}), albeit with
larger exponents $\alpha_i$. For that reason, all these terms make
progressively less visible (but not necessarily unobservable)
contributions to the spectral function.


We thank Matt Grayson, Woowon Kang, Leon Balents and Lloyd Engel for
helpful discussions. This work was supported by NSF grant No. DMR-0225698.


\end{document}